\documentclass[12pt,a4paper]{article}

\usepackage{epsfig}
\usepackage{psfig}
\def\lsim{\raise0.3ex\hbox{$<$\kern-0.75em\raise-1.1ex\hbox{$\sim$}}}
\def\gsim{\raise0.3ex\hbox{$>$\kern-0.75em\raise-1.1ex\hbox{$\sim$}}}
\newcommand{\be}{\begin{equation}}
\newcommand{\ee}{\end{equation}}
\newcommand{\ba}{\begin{eqnarray}}
\newcommand{\ea}{\end{eqnarray}}

\def\spose#1{\hbox to 0pt{#1\hss}}
\def\ltapprox{\mathrel{\spose{\lower 3pt\hbox{$\mathchar"218$}}
 \raise 2.0pt\hbox{$\mathchar"13C$}}}
\def\gtapprox{\mathrel{\spose{\lower 3pt\hbox{$\mathchar"218$}}
 \raise 2.0pt\hbox{$\mathchar"13E$}}}

\def\ad#1{$\,^{\rm #1}$}
\def\NT{N_\tau}
\def\nt{\ifmmode\NT\else$\NT$\fi}
\def\NS{N_\sigma}
\def\ns{\ifmmode\NS\else$\NS$\fi}

\def\PR{{ Phys.\ Rev.\ }}

\def\PRL{{ Phys.\ Rev.\ Lett.\ }}
\def\PL{{ Phys.\ Lett.\ }}
\def\NP{{ Nucl.\ Phys.\ }}

\def\PBP{{\langle \bar\psi\psi \rangle }}

\def\p{^\prime}

\def\n{\noindent}

\begin{document}
\begin{titlepage} 
\thispagestyle{empty}

 \mbox{} \hfill BI-TP 2001/07\\
 \mbox{} \hfill May 2001
\begin{center}
\vspace*{1.0cm}
{{\Large \bf Finite-size-scaling functions for  \\         
  $3d$ $O(4)$ and $O(2)$ spin models and QCD\\}}\vspace*{1.0cm}
{\large J. Engels\ad a, S. Holtmann\ad a, T. Mendes\ad b and
 T. Schulze\ad a}\\ \vspace*{0.8cm}
\centerline {{\large $^{\rm a}$}{\em Fakult\"at f\"ur Physik, 
    Universit\"at Bielefeld, D-33615 Bielefeld, Germany}} \vspace*{0.4cm}
\centerline {{\large $^{\rm b}$}{\em IFSC-USP, Caixa postal 369,
    13560-970 S\~ao Carlos SP, Brazil}} \vspace*{0.4cm}
\protect\date \\ \vspace*{0.9cm}
{\bf   Abstract   \\ } \end{center} \indent
We calculate numerically universal finite-size-scaling functions for
the three-dimen\-sio\-nal $O(4)$ and $O(2)$ models. The approach of these
functions to the infinite-volume scaling functions is studied in 
detail on the critical and pseudocritical lines. For this purpose we
determine the pseudocritical line in two different ways. We find that 
the asymptotic form of the finite-size-scaling functions is already 
reached at small values of the scaling variable. A comparison with QCD 
lattice data for two flavours of staggered fermions shows a similar 
finite-size behaviour which {\em is compatible} with that of the 
spin models.
\vfill \begin{flushleft} 
\noindent{\rule[-.3cm]{5cm}{.02cm}} \\
\vspace*{0.2cm} 
E-mail: engels,holtmann,tschulze@physik.uni-bielefeld.de; mendes@if.sc.usp.br
\end{flushleft} 
\end{titlepage}


\section{Introduction}

At finite temperature quantum chromodynamics (QCD) undergoes a chiral phase
transition. For two degenerate light-quark flavours this transition is 
supposed to be of 
second order in the continuum limit and to belong to the same universality 
class as the $3d$ $O(4)$ model \cite{PW}-\cite{RW}. QCD lattice data have 
therefore been compared to the universal $O(4)$ scaling function 
\cite{first}-\cite{Ejiri}. The scaling function or equation of state describes
the system in the thermodynamic limit, that is for $V\rightarrow \infty$.
It was first determined numerically in Ref.\ \cite{Toussaint} and later
studied in more detail in Ref.\ \cite{EM}. Lattice results for Wilson 
fermions \cite{Iwas,wilson} seem to agree quite well with the predictions,
though for the Wilson action the chiral symmetry is only restored in the 
continuum limit. In the staggered formulation of QCD a part of the 
chiral symmetry is remaining even for finite lattice spacing, and that
is $O(2)$. Nevertheless, comparisons with $O(4)$ or even $O(2)$ scaling 
functions \cite{o2} have up to now not confirmed the expectations for
staggered fermions \cite{JLQCD}-\cite{MILC}. Among the many arguments 
\cite{MILC,Ber}, which have been put forward to explain this failure,
one is obvious, namely, that lattice QCD simulations are still performed
on relatively small volumes and therefore will show substantial
finite size effects. More adequate tests may be carried out, if
universal finite-size-scaling functions for the $O(N)$-spin models are
available. This exactly is the aim of the paper: the  calculation of 
finite-size-scaling functions for the $O(4)$ and $O(2)$ models and a 
corresponding test of QCD lattice data. 

We shall make extensive use of the results of two of our papers: a study 
of the three-dimensional $O(4)$ model, Ref.\ \cite{EM}, and another one
for the $O(2)$ model, Ref.\ \cite{o2}. There we determined the respective
equations of state. In the following we briefly review the equations
which are relevant for this paper.

The $O(N)$-invariant nonlinear $\sigma$-models, which we 
investigate are defined by
\be
\beta\,{\cal H}\;=\;-J \,\sum_{<i,j>} {\bf S}_i\cdot {\bf S}_j
         \;-\; {\bf H}\cdot\,\sum_{i} {\bf S}_i \;,
\ee
where $i$ and $j$ are nearest-neighbour sites on a $d-$dimensional 
hypercubic lattice, and ${\bf S}_i$ is an $N$-component unit vector 
at site $i$. It is convenient to decompose 
the spin vector ${\bf S}_i$ into a longitudinal (parallel to the magnetic 
field ${\bf H}$) and a transverse component 
\be
{\bf S}_i\; =\; S_i^{\parallel} {\bf \hat H} + {\bf S}_i^{\perp} ~.
\ee
The order parameter of the system, the magnetization $M$, is then the 
expectation value of the lattice average $S^{\parallel}$
of the longitudinal spin component
\be
M \;=\; <\!\frac{1}{V}\sum_{i} S_i^{\parallel}>\; =\; <  S^{\parallel}>~.
\ee
There are two types of susceptibilities: the longitudinal 
susceptibility is defined as usual by the derivative of the magnetization, 
whereas the transverse susceptibility corresponds to the fluctuation 
of the lattice average ${\bf S}^{\perp}$ of the transverse spin per component
\ba
\chi_L\!\! &\!=\!&\!\! {\partial M \over \partial H}
 \;=\; V(<{ S^{\parallel}}^2>-M^2)~, \label{chil}\\
\chi_T\!\! &\!=\!&\!\!{V \over N-1}< {{\bf S}^{\perp}}^2> 
\;=\; {M \over H}
~. \label{chit}
\ea
We do not discuss here as in \cite{EM} and \cite{o2} the singularities of 
the susceptibilities on the coexistence line which are
due to the Goldstone modes. We simply note, that 
the general Widom-Griffiths form of the equation of state \cite{Griffiths},
which describes the critical behaviour of the magnetization
in the vicinity of $T_c$, is compatible with these singularities. It
is given by
\be
y\;=\;f(x)\;,
\label{eqstate}
\ee
where 
\be
y \equiv h/M^{\delta}, \quad x \equiv t/M^{1/\beta}.
\label{xy}
\ee
The variables $t$ and $h$ are the normalized 
reduced temperature $t=(T-T_c)/T_0$ and magnetic field $h=H/H_0$.
We take the usual normalization conditions 
\be
f(0) = 1, \quad f(-1) = 0~.
\label{normal}
\ee
The critical exponents $\delta$ and $\beta$ appearing in Eqs.~\ref{eqstate}
and \ref{xy} specify all the other critical exponents
\be
d\nu=\beta(1+\delta),\quad\gamma=\beta(\delta-1),\quad \nu_c=\nu/\beta\delta~.
\ee
Possible irrelevant scaling fields and exponents are however not taken 
into account in Eq.\ \ref{eqstate}, the function $f(x)$ 
is universal. Another way to express the dependence of the magnetization 
on $t$ and $h$ is
\be
M\;=\;h^{1/\delta} f_G(t/h^{1/\beta\delta})~,
\label{ftous}
\ee
where $f_G$ is a scaling function. This type of 
scaling equation is used for comparison to QCD lattice data.
The scaling forms in Eqs.\ (\ref{eqstate}) and (\ref{ftous}) are 
clearly equivalent, since the variables $x$ and $y$ are related 
to the scaling function $f_G$ and its argument by
\be
 y\;=\;f_G^{-\delta} \;, \quad  
x \;=\;(t/h^{1/\beta\delta})\, f_G^{-1/\beta}\;.
\ee
In Refs. \cite{EM} and \cite{o2} we had parametrized the equation of
state by a combination of a small-$x$ (low temperature) form $x_s(y)$,
which was inspired by the approximation of Wallace and Zia \cite{WZ} close
to the coexistence line ($x=-1;~y=0$)
\be
x_s(y)+1 \;=\; ({\widetilde c_1} \,+\, {\widetilde d_3})\,y \,+\,
             {\widetilde c_2}\,y^{1/2} \,+\, 
             {\widetilde d_2}\,y^{3/2} \;,
\label{PTform}
\ee
and a large-$x$ (high temperature) form $x_l(y)$ derived from  
Griffiths's analyticity condition\cite{Griffiths}
\be
x_l(y)\;=\; a\, y^{1/\gamma} \,+\, b\,y^{(1-2\beta)/\gamma}~.
\label{highx}
\ee
\begin{table}[ht]
\begin{center}
  \begin{tabular}{|ccc||cc||cc||c|}
    \hline
${\widetilde c_1} \,+\, {\widetilde d_3}$ &${\widetilde c_2}$ &
$ {\widetilde d_2}$ & $a$ & $b$ & $y_0$ & $p$& \\ \hline \hline
 0.345(12)& 0.674(08)& -0.023(5)& 1.084(6)& -0.994(109)
& 10.0 & 3 & $O(4)$ \\ \hline                 
 0.352(30)& 0.592(10) & 0.056 & 1.260(3)& -1.163(20)
& $\;~3.5$ & 6 & $O(2)$ \\ \hline                 
  \end{tabular}
\end{center}
\caption{Parameters of the fits to the scaling functions for $O(4)$
and $O(2)$.}

\label{tab:param}
\end{table}
\begin{table}[ht]
\begin{center}
  \begin{tabular}{|ccccc||ccc||c|}
    \hline
$\beta$ &$\delta$ & $\gamma$ & $\nu$ & $\nu_c$ 
& $J_c$ & $T_0$ & $H_0$ & \\ \hline \hline
 0.380 & 4.86 & 1.4668 & 0.7423 & 0.4019
& 0.93590 & 1.093 & 5.08 & $O(4)$ \\ \hline                 
 0.349 & 4.7798 & 1.3192 & 0.6724 & 0.4031 
& 0.454165 & 1.18 & 1.11 & $O(2)$ \\ \hline                 
  \end{tabular}
\end{center}
\caption{Critical parameters used in the $O(4)$ \cite{EM} and $O(2)$ \cite{o2}
calculations.}

\label{tab:critic}
\end{table}
\n The two parts can be interpolated smoothly by an ansatz of the kind
\be
x(y) \;=\; x_s(y)\,\frac{y_0^p}{y_0^p + y^p} \,+\,
           x_l(y)\,\frac{y^p}{y_0^p + y^p}~,
\label{totalfit}
\ee
from which the total scaling function is obtained. In Table
\ref{tab:param} the parameters of these fits are listed. Two remarks
are necessary here: for $O(2)$ the coefficient ${\widetilde d_2}$ was 
fixed by the normalization $y(0)=1$, that is ${\widetilde d_2}= 1-(   
{\widetilde c_1}+{\widetilde d_3}+{\widetilde c_2})$, and in the $O(4)$
case the coefficient $b$ was incorrectly cited in Ref.\ \cite{EM}. Of course,
the scaling functions are not independent of the critical points, amplitudes
and exponents, which had been used in their determination. For completeness
we therefore give in Table \ref{tab:critic} the relevant input.

\section{Finite-Size-Scaling Functions}
\label{section:FSSF}

The general form of the finite-size-scaling function for the magnetization
is given by
\be
M = L^{-\beta/\nu} \Phi ( tL^{1/\nu}, hL^{1/\nu_c}, L^{-\omega})~,
\label{fssm}
\ee
that is, we have a function of three (or even more) variables, which describes
the dependence of the magnetization on the thermal, magnetic and possible
irrelevant scaling fields and the characteristic linear extension $L$ of the
volume. Here we have specified only the leading irrelevant 
scaling field proportional to $L^{-\omega}$, with $\omega >0$. A universal 
scaling function is obtained, when we expand the function $\Phi$ in 
$L^{-\omega}$ and consider the first term only 
\be
M = L^{-\beta/\nu} \Phi_0 ( tL^{1/\nu}, hL^{1/\nu_c})  + \dots~.
\label{mexpan}
\ee
\begin{figure}[htb]
\begin{center}
   \epsfig{bbllx=127,bblly=264,bburx=451,bbury=587,
        file=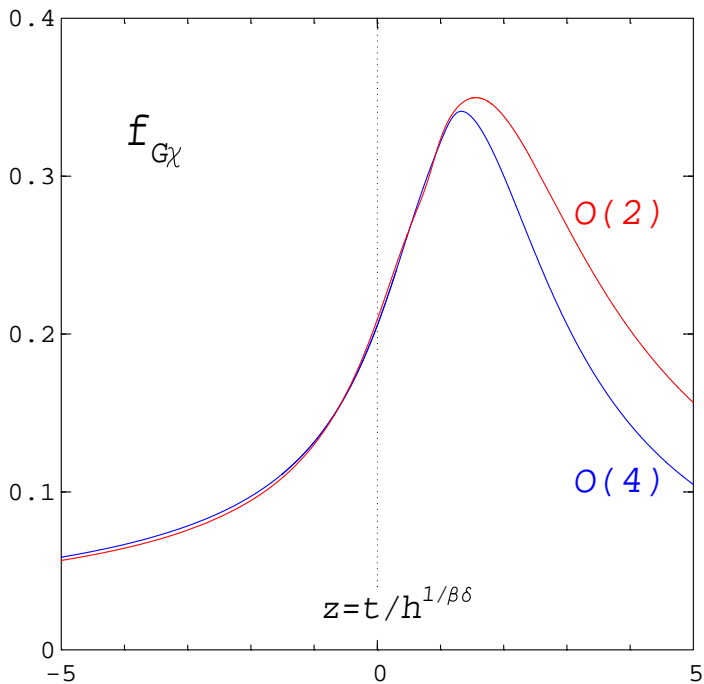,width=84mm}
\end{center}
\caption{The scaling function $f_{G\chi}(z)$ of the longitudinal
susceptibility for the $O(4)$ and $O(2)$ models.} 
\label{fig:chil}
\end{figure}
The function $\Phi_0$ still depends on two variables.
In order to handle the two-variable dependence of $\Phi_0$ in an economic
way, we consider in the following paths in the $(t,h)$-plane defined by
fixed values of $z=th^{-1/\beta\delta}$. At fixed $z$ we can express one
of the two variables of Eq.\ (\ref{mexpan}) by $z$ and the other variable,
leaving us with a function of one variable only
\be
M = L^{-\beta/\nu} Q_z ( hL^{1/\nu_c})  + \dots~,
\label{mq0}
\ee
where $Q_z$ is again universal.
The procedure has the additional advantage that $z$ is the argument of
the scaling function $f_G$ of Eq.\ (\ref{ftous}), thus requiring only 
one point of $f_G$ to calculate the asymptotic form $Q_{z,\infty}$
of the finite-size-scaling function $Q_z$
\be
Q_z \rightarrow  Q_{z,\infty} = f_G(z)( hL^{1/\nu_c})^{1/\delta}
\quad \mbox{for}\quad L\rightarrow \infty~.
\label{Qasy}
\ee
Examples of lines of fixed $z$
are the critical line where $z=0$ and the pseudocritical line, the line of
peak positions of the susceptibility $\chi_L$ in the $(t,h)$-plane
for $V\rightarrow \infty$. There are two ways to find that value of $z$ for
$O(N)$, which corresponds to the pseudocritcal line. One way amounts to
locating the peak positions of $\chi_L$ as a function of the temperature at 
different fixed small values of the magnetic field on lattices with increasing 
size $L^3$. This method has been used in QCD. For staggered fermions the 
pseudocritical line thus found shows up to now the most convincing agreement 
with the $O(N)$ models. The scaling function offers a more elegant way 
to determine the pseudocritical line. Since $\chi_L$ is the derivative of $M$ 
\be
\chi_L={\partial M\over \partial H}={h^{1/\delta-1} \over H_0} f_{G\chi}(z)~,
\label{max}
\ee
its scaling function $f_{G\chi}(z)$ can be calculated directly from $f_G(z)$
\be
f_{G\chi}(z)={1 \over \delta}
\left ( f_G(z) - {z\over \beta}f_G\p(z) \right)~.
\label{fgc}
\ee
Evidently, the maximum of $\chi_L$ at fixed $h$ and varying $t$ is at 
the peak position $z_p$ of  $f_{G\chi}(z)$ and  $z_p$ is another universal
quantity. In Fig.\ \ref{fig:chil} we show the result for $f_{G\chi}(z)$ from 
Eq.\ (\ref{fgc}) using the scaling functions $f_G$ for $O(4)$ and $O(2)$ 
as obtained from Table \ref{tab:param}. In this calculation we have 
interpolated the small-$x$ and large-$x$ derivatives $y(dx/dy)$ to smooth the
result. We see from Fig.\ \ref{fig:chil} that there is a relatively broad
peak at positive $z$ and we can read off the value of $z_p$ for the two
models. They are listed in Table \ref{tab:zp} together with the peak height
of $f_{G\chi}$. It is instructive to use the 
\begin{table}[h!]
\begin{center}
  \begin{tabular}{|cc||c||c|}
    \hline
\multicolumn{2}{|c||}{Scaling function} & $L=24$ to 96 & \\ \cline{1-3} 
$z_p$ & $f_{G\chi}(z_p)$ & $z_p$ & \\ \hline \hline
$1.33 \pm 0.05$ & 0.341(1)  & $1.35 \pm 0.10$ & $O(4)$ \\ \hline                 
$1.56 \pm 0.10$ & 0.350(1)  & $1.65 \pm 0.10$ & $O(2)$ \\ \hline                 
  \end{tabular}
\end{center}
\caption{The peak position and height of the scaling function $f_{G\chi}\,$, 
and $z_p$ from calculations on lattices with size $L=24$ to 96.}

\label{tab:zp}
\end{table}

\n QCD method for $z_p$ determination on finite 
lattices as well. For that purpose we have calculated at eight values of the 
magnetic field on lattices of size $L=24,\dots,96$ the peak positions and heights 
for $O(4)$. The infinite volume estimates for the two quantities are compared
in Fig.\ \ref{fig:poshei} and Table \ref{tab:zp} to the results from the scaling 
function. We observe in Fig.\ \ref{fig:poshei}a that the agreement is very
good for the peak positions at small $h$. At larger $h$ there is a slight
tendency towards somewhat higher pseudocritical temperatures than expected
from the fixed $z$ relation between $t$ and $h$ at the peak. The peak heights 
in Fig.\ \ref{fig:poshei}b on the other hand are following nicely the
prediction $\chi_L=0.244 H^{1/\delta -1}$ from the scaling function for all
$H$. We have obtained similar results for $O(2)$ at two values of $H$. Fig.\
\ref{fig:poshei}a contains also lines for several other fixed $z$ values 
to give a better overview of the $(t,h)$-plane. As examples we shall 
investigate in the next two subsections the finite-size behaviour of the 
magnetization on the lines $z=0$ and $z=z_p$.

\subsection{Finite-Size Scaling in the $O(4)$ Model}

\n Our simulations are done on three-dimensional lattices with periodic
boundary conditions and linear extensions $L$ up to 120. We use the same
cluster algorithm as in Refs.\ \cite{EM,o2}. Let us first consider the 
critical line in $O(4)$. In Fig.\ 4b of Ref.\ \cite{EM} we had 
observed, that there are essentially no corrections to scaling on the 
critical line. Here we extend this investigation by including more 
points at higher and also very
\newpage
\setlength{\unitlength}{1cm}
\begin{picture}(13,7.3)
\put(0,0){
   \epsfig{bbllx=127,bblly=264,bburx=451,bbury=587,
       file=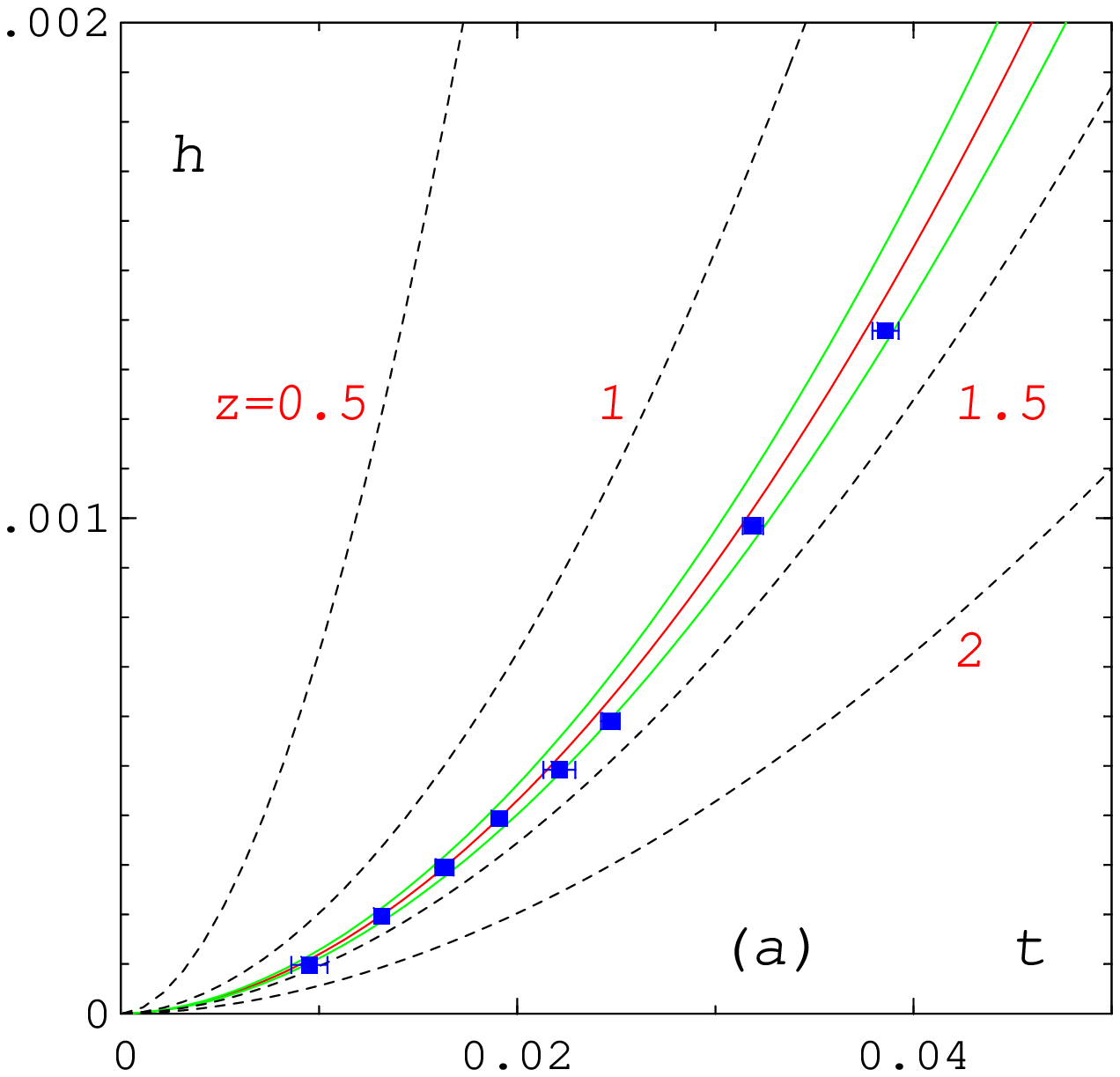, width=67mm}
          }
\put(7.5,0){
   \epsfig{bbllx=127,bblly=264,bburx=451,bbury=587,
       file=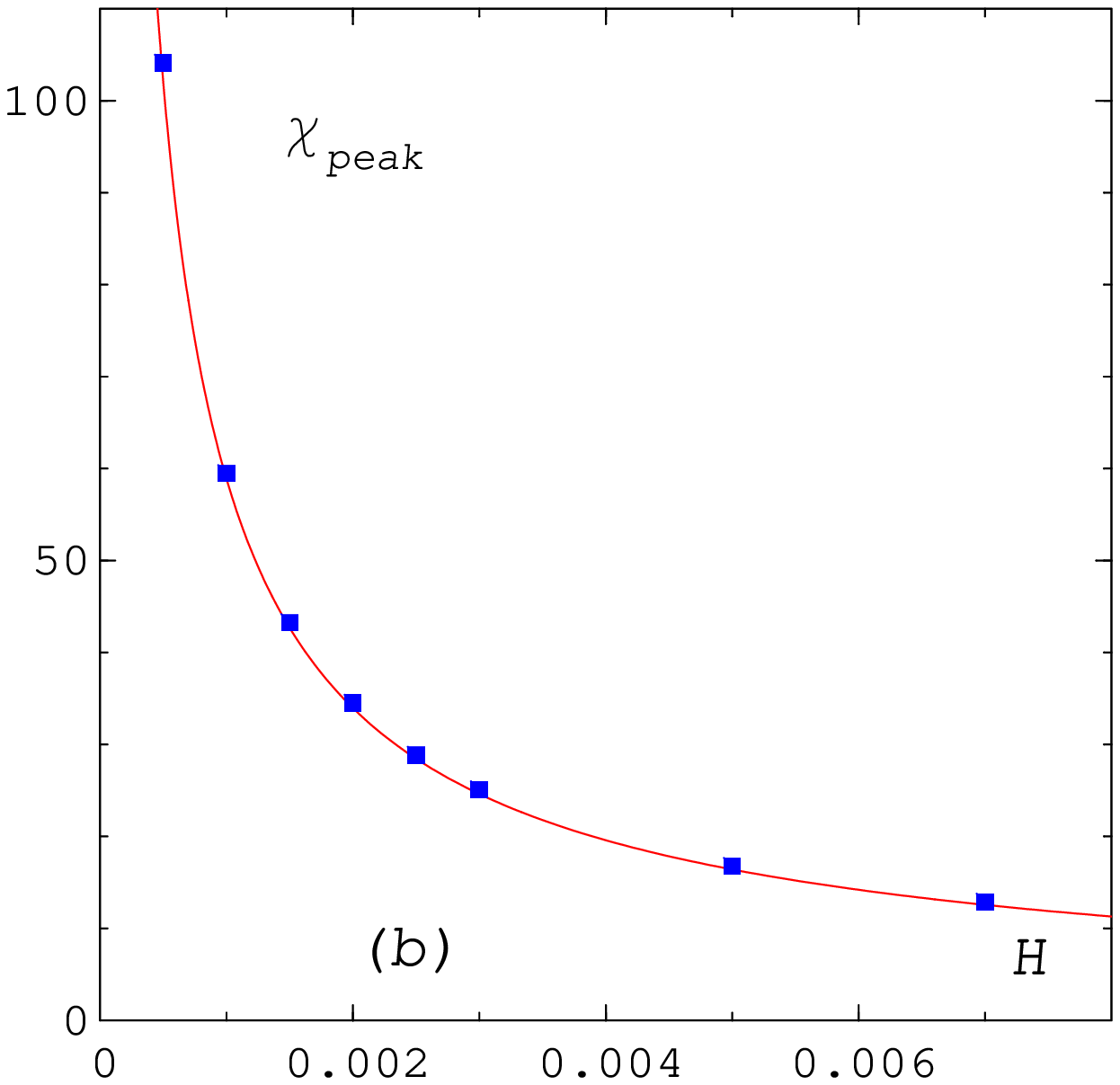, width=67mm}
          }
\end{picture}
\begin{figure}[h!]
\caption{(a) Lines of fixed $z=0.5,1,1.5,2$ (dashes), the pseudocritical
line (solid) at $z_p=1.33\pm 0.05$ and measured peak positions
(squares). (b) the peak height of $\chi_L$ as a function of $H$, measured
(squares) and from the scaling funtion (solid line).
Both parts of the figure refer to the $O(4)$ model.}
\label{fig:poshei}
\end{figure}

\setlength{\unitlength}{1cm}
\begin{picture}(13,6.5)
\put(0,0){
   \epsfig{bbllx=127,bblly=264,bburx=451,bbury=587,
       file=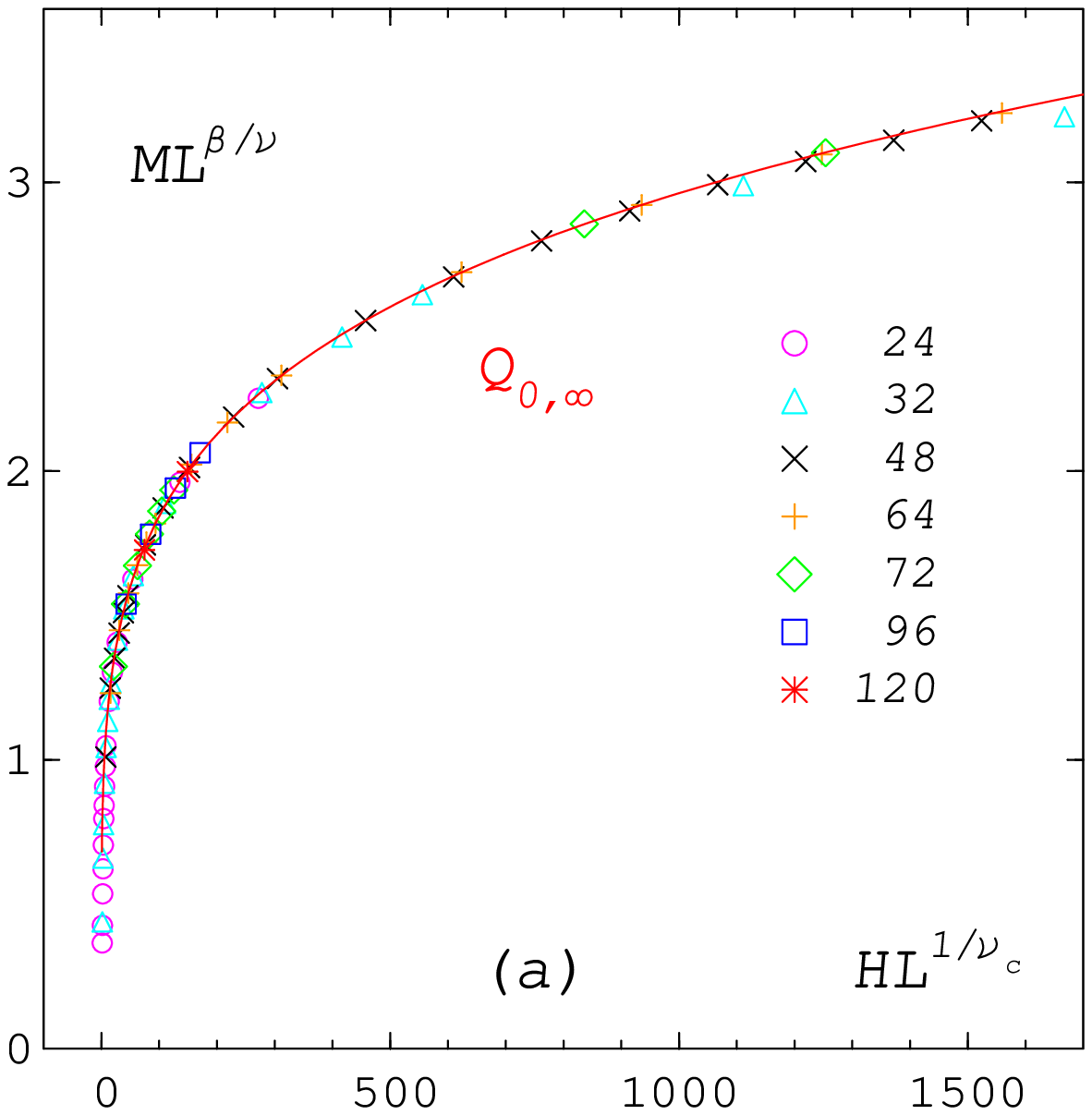, width=67mm}
          }
\put(7.5,0){
   \epsfig{bbllx=127,bblly=264,bburx=451,bbury=587,
       file=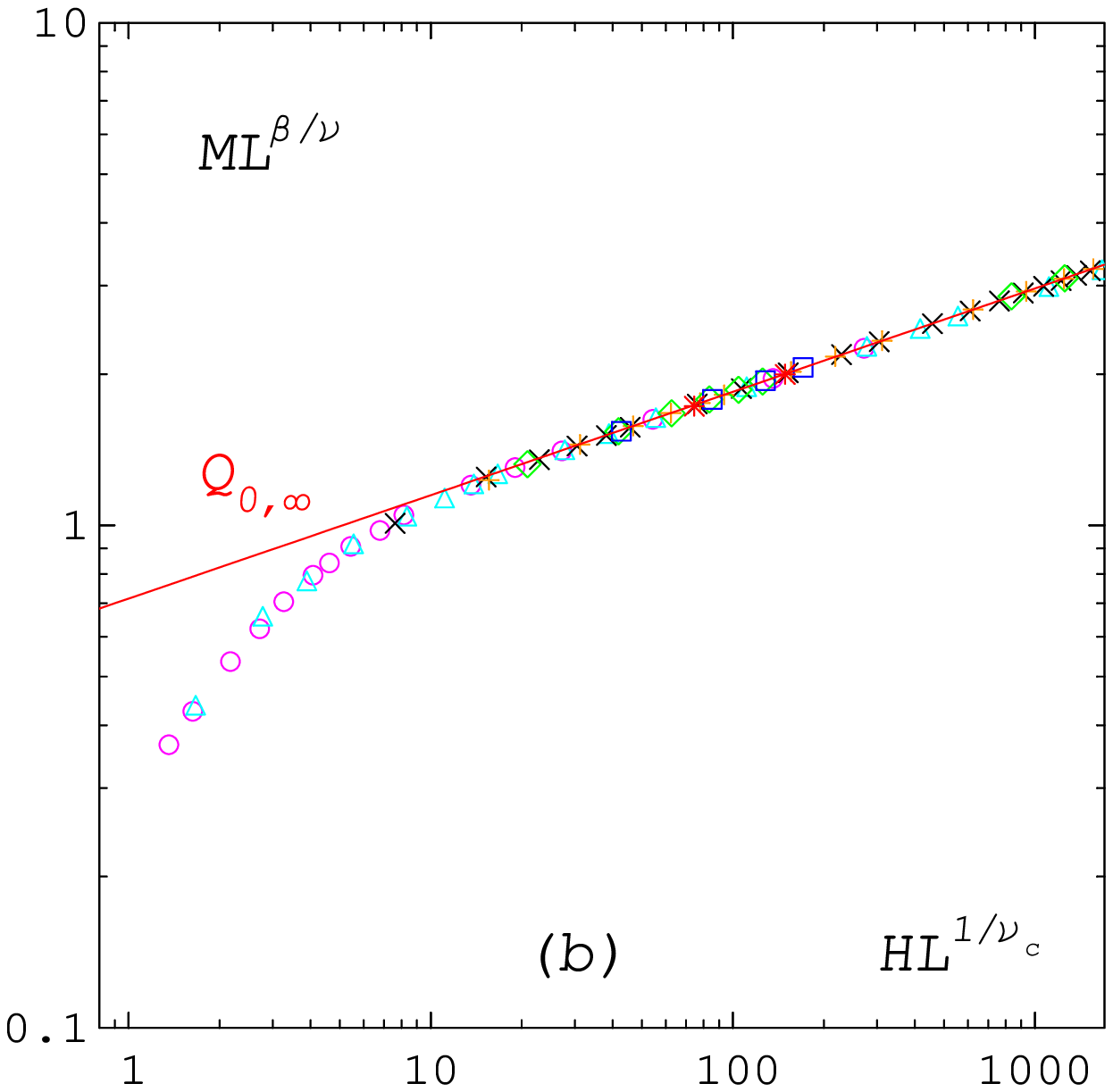, width=67mm}
          }
\end{picture}
\begin{figure}[h!]
\caption{(a) Finite-size scaling of $ML^{\beta/\nu}$ for $O(4)$, Eq.\ 
\ref{mq0}, on the critical line. The solid line shows the
asymptotic form $Q_{0,\infty}$, the symbols denote different lattice
sizes $L$. (b) is a double-log plot of (a).}
\label{fig:o40}
\end{figure}

\n 
small values of the scaling variable $HL^{1/\nu_c}$. The new scaling plot 
is shown in Fig.\ \ref{fig:o40}a. With the higher amount of data we find 
that the finite-size-scaling function $Q_0$ is actually reached from below
with increasing $L$, though the differences between different $L$ are 
hardly visible. In Fig.\ \ref{fig:o40}b, where we show the same data
logarithmically, we see that $Q_0$ approaches $Q_{0,\infty}$ from below
and coincides with its asymptotic form already at about $HL^{1/\nu_c}
\approx 20$. 

\n We have also calculated the magnetization on the pseudocritical line
(for $O(4)$ at $z_p=1.33$) on a variety of finite lattices. The scaling plot
is shown in Fig.\ \ref{fig:o4pc}a. It differs from Fig.\ \ref{fig:o40}a 
in several respects. There are strong corrections to scaling and the
approach to the universal function $Q_{z_p}$ is from above. If one looks at
the logarithmic plot, Fig.\ \ref{fig:o4pc}b, one finds a similar increase
at small $HL^{1/\nu_c}$ as in the case of the critical line. Here the
asymptotic form $Q_{z_p,\infty}$ is reached around $HL^{1/\nu_c}\approx 30$.
Since $Q_{z_p,\infty}$ was calculated from Eq.\ \ref{Qasy} we confirm 
herewith also the value of $f_G(z_p)$.
\setlength{\unitlength}{1cm}
\begin{picture}(13,7.2)
\put(0.5,0){
   \epsfig{bbllx=127,bblly=264,bburx=451,bbury=587,
       file=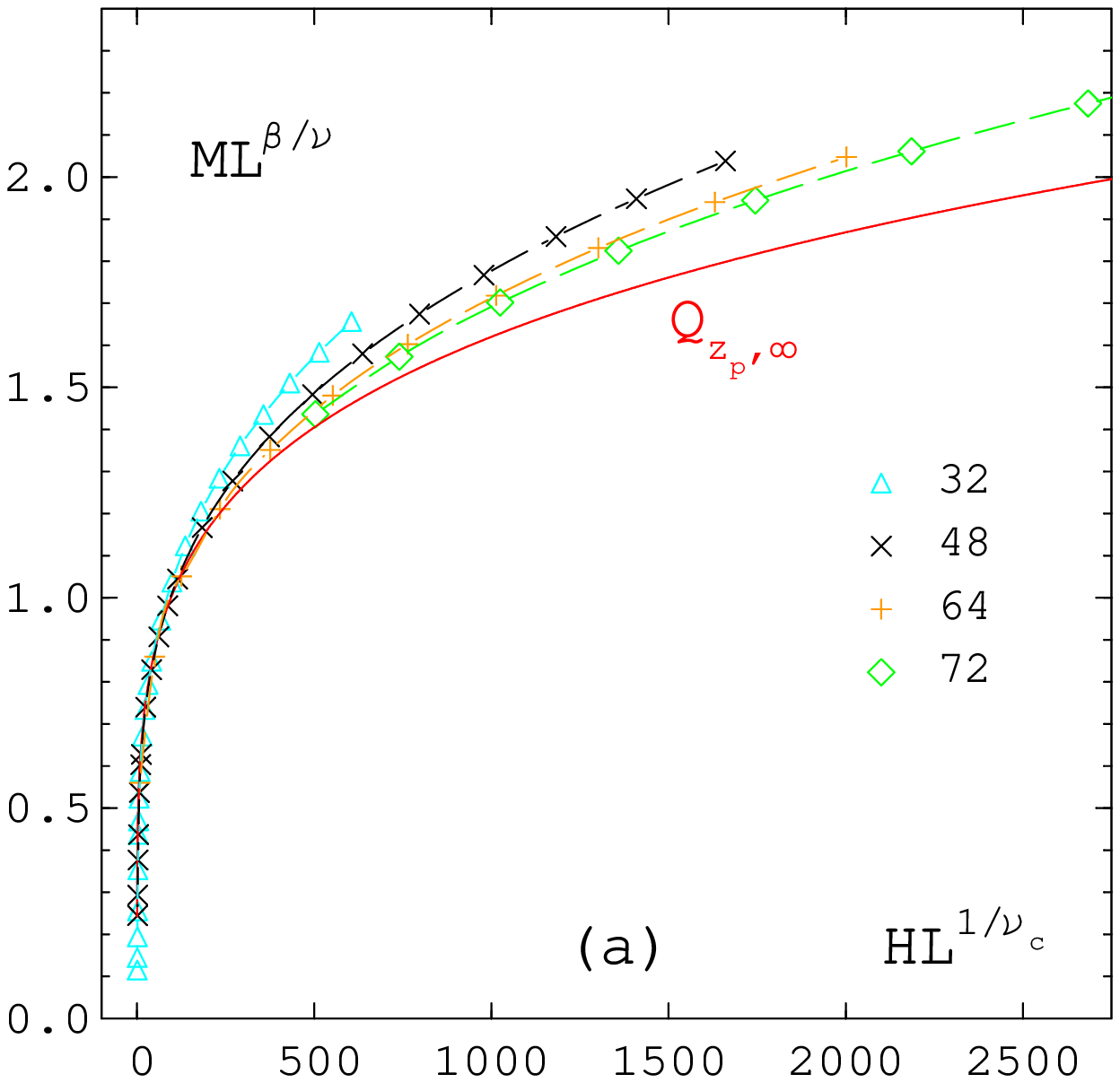, width=67mm}
          }
\put(8.0,0){
   \epsfig{bbllx=127,bblly=264,bburx=451,bbury=587,
       file=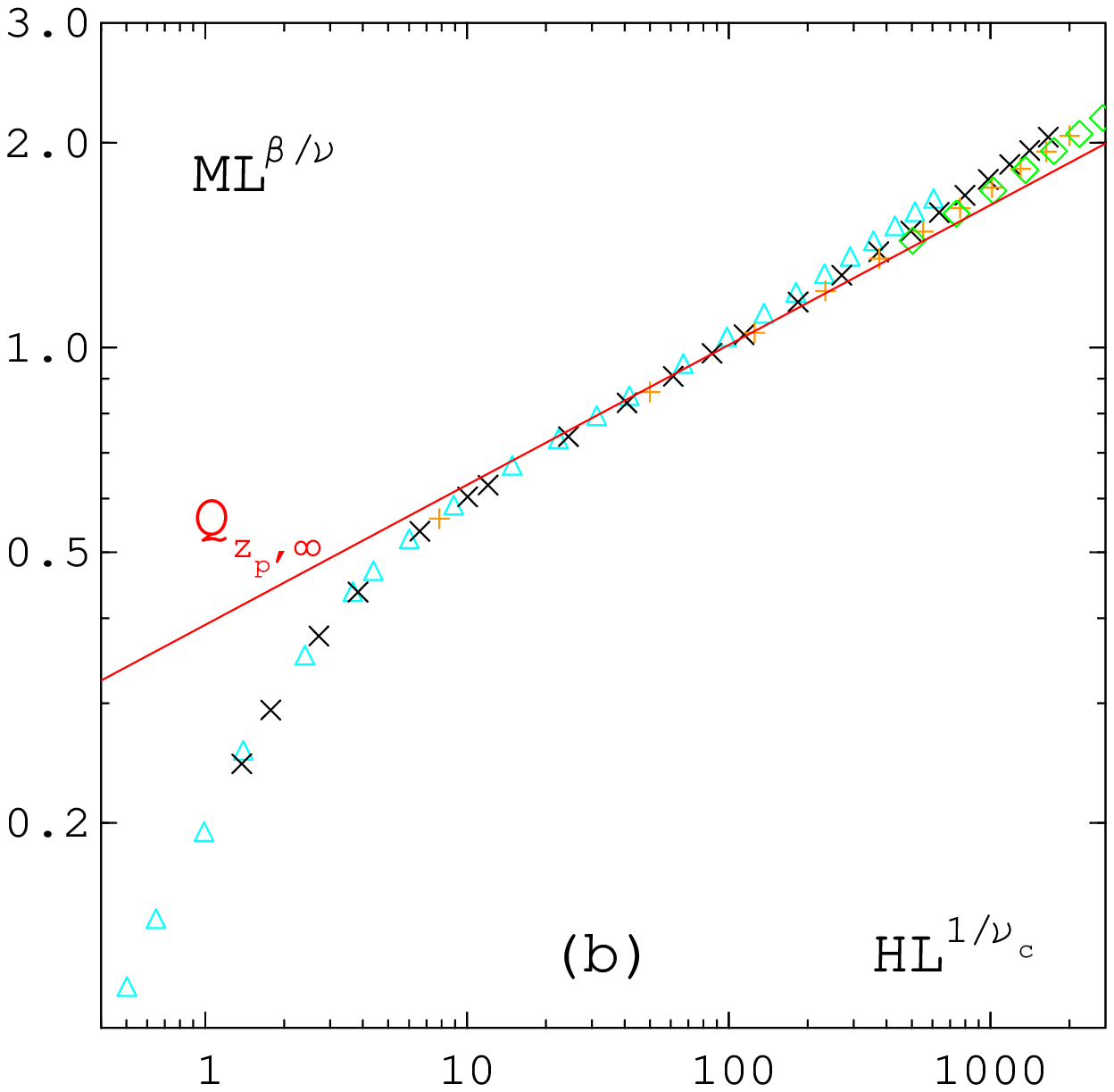, width=67mm}
          }
\end{picture}
\begin{figure}[h!]
\caption{(a) Finite-size scaling of $ML^{\beta/\nu}$ for $O(4)$
on the pseudocritical line. The solid line shows the
asymptotic form $Q_{z_p,\infty}$, the symbols denote different lattice
sizes $L$. (b) is a double-log plot of (a).}
\label{fig:o4pc}
\end{figure}

\subsection{Finite-Size Scaling in the $O(2)$ Model}

In Ref.\ \cite{o2} we have found negative corrections to scaling on the 
coexistence line and less pronounced ones also on the critical line of the 
$O(2)$ model in the thermodynamic limit. The occurrence of these corrections
is well understood by renormalization-group theory \cite{Bagnuls}. On finite 
lattices we expect because of the corrections considerable finite-size 
effects on the critical line. We have calculated the magnetization on 8 
lattices with $L=8$ to 96 \cite{latt} and show the results from the
reweighted data in Fig.\ \ref{fig:o20}a. From these curves we have estimated
the universal scaling function $Q_0$ by square fits in $L^{-\omega}$ at fixed
values of $HL^{1/\nu_c}$. The exponent ${\omega}=0.79(2)$ was taken from 
Ref.\ \cite{Hase}. In Fig.\ \ref{fig:o20}b we compare $Q_0$ to the asymptotic
form $Q_{0,\infty}$ and data for
\newpage
\setlength{\unitlength}{1cm}
\begin{picture}(13,7.2)
\put(-0.15,0){
   \epsfig{bbllx=127,bblly=264,bburx=451,bbury=587,
       file=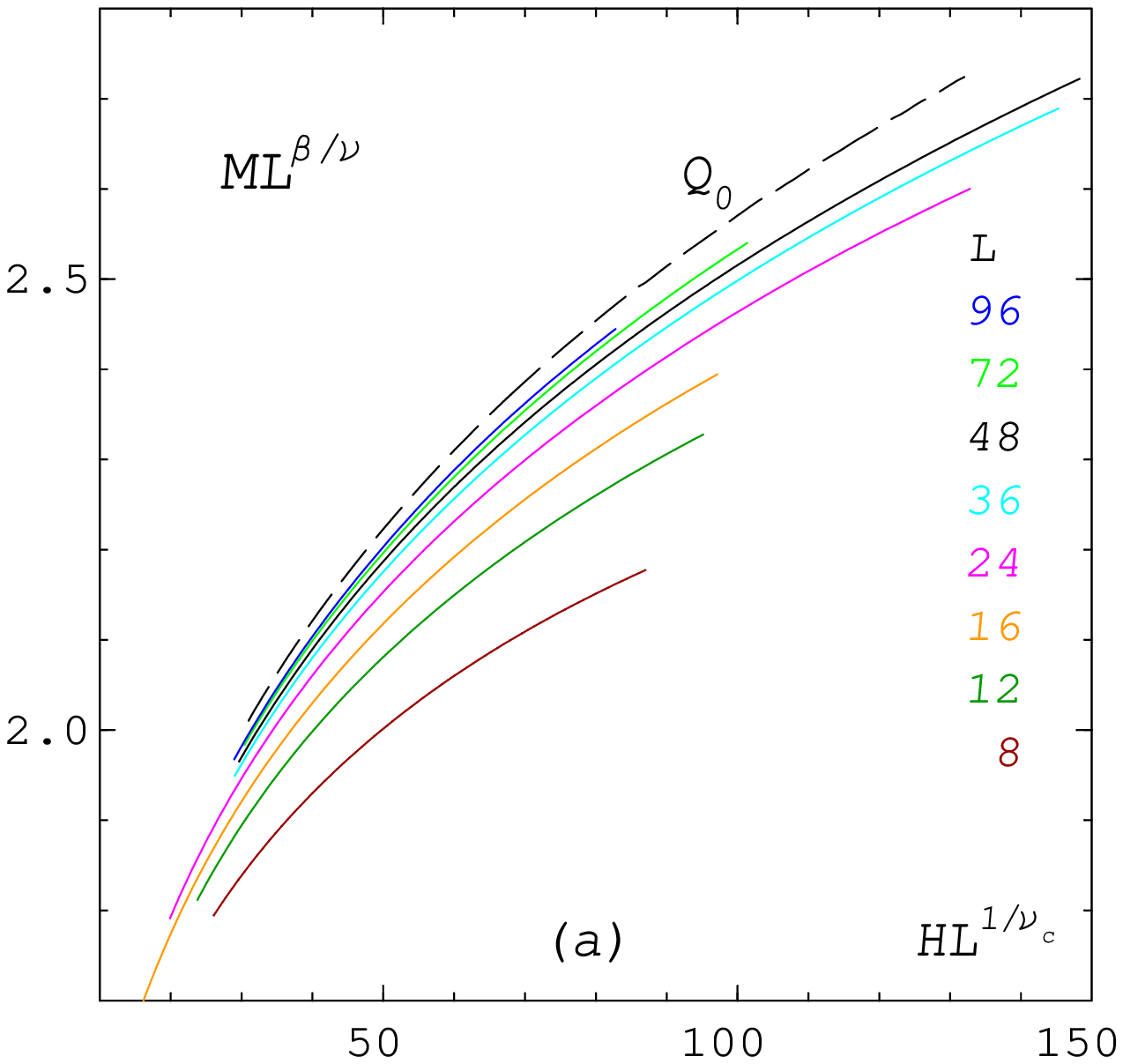, width=67mm}
          }
\put(7.35,0){
   \epsfig{bbllx=127,bblly=264,bburx=451,bbury=587,
       file=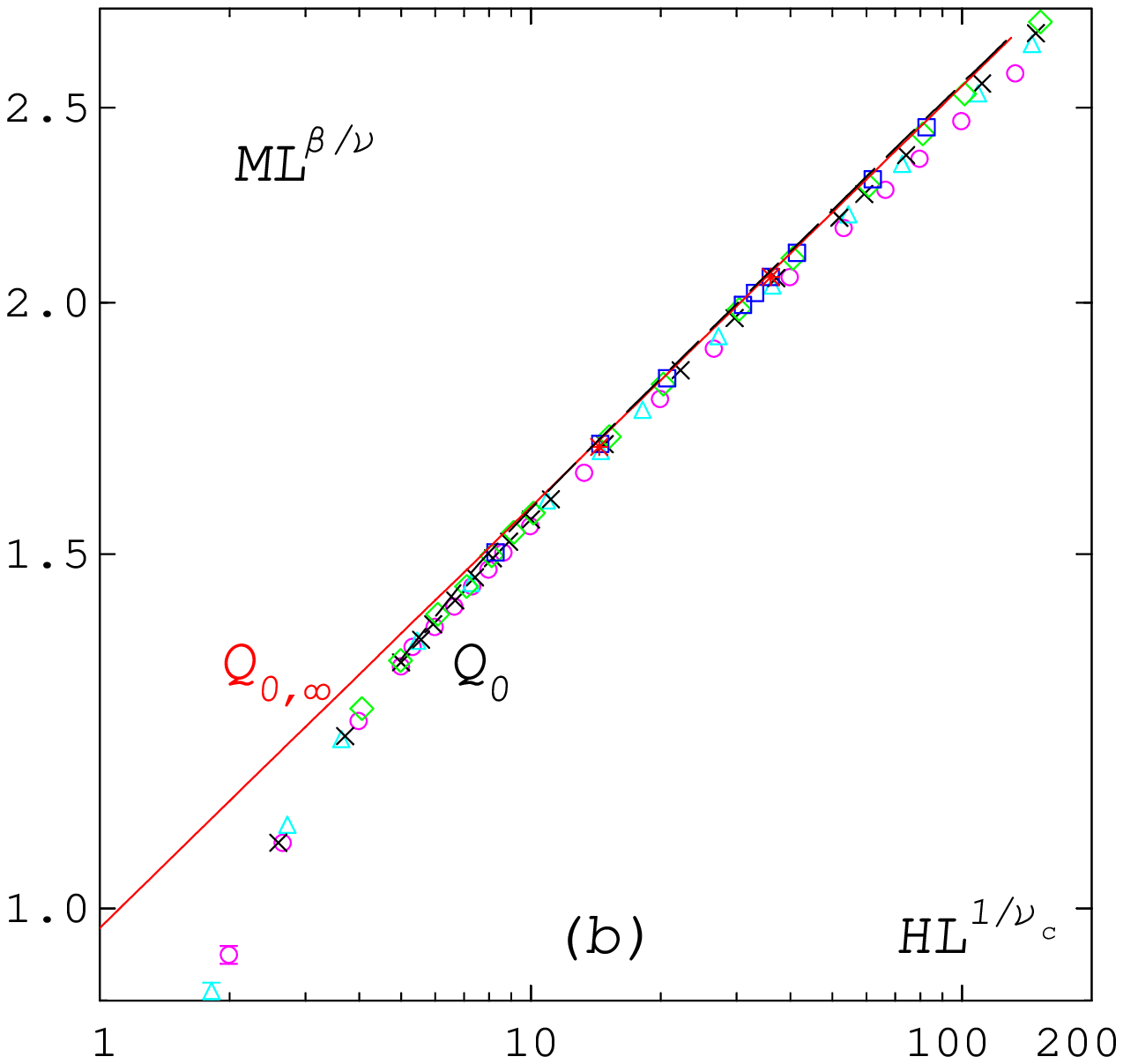, width=67mm}
          }
\end{picture}
\begin{figure}[h!]
\caption{(a) $ML^{\beta/\nu}$ on the critical line for $O(2)$ from reweighted
data (solid lines) on lattices with different $L$. The dashed line shows
the estimate for $Q_0$. (b) is a double-log plot of (a), including the
asymptotic form $Q_{0,\infty}$ (solid line) and  with the direct data 
for $L \ge 24$ (notation like in Fig.\ \ref{fig:o40}a) replacing the 
reweighting lines.}
\label{fig:o20}
\end{figure}

\setlength{\unitlength}{1cm}
\begin{picture}(13,6.5)
\put(-0.15,0){
   \epsfig{bbllx=127,bblly=264,bburx=451,bbury=587,
       file=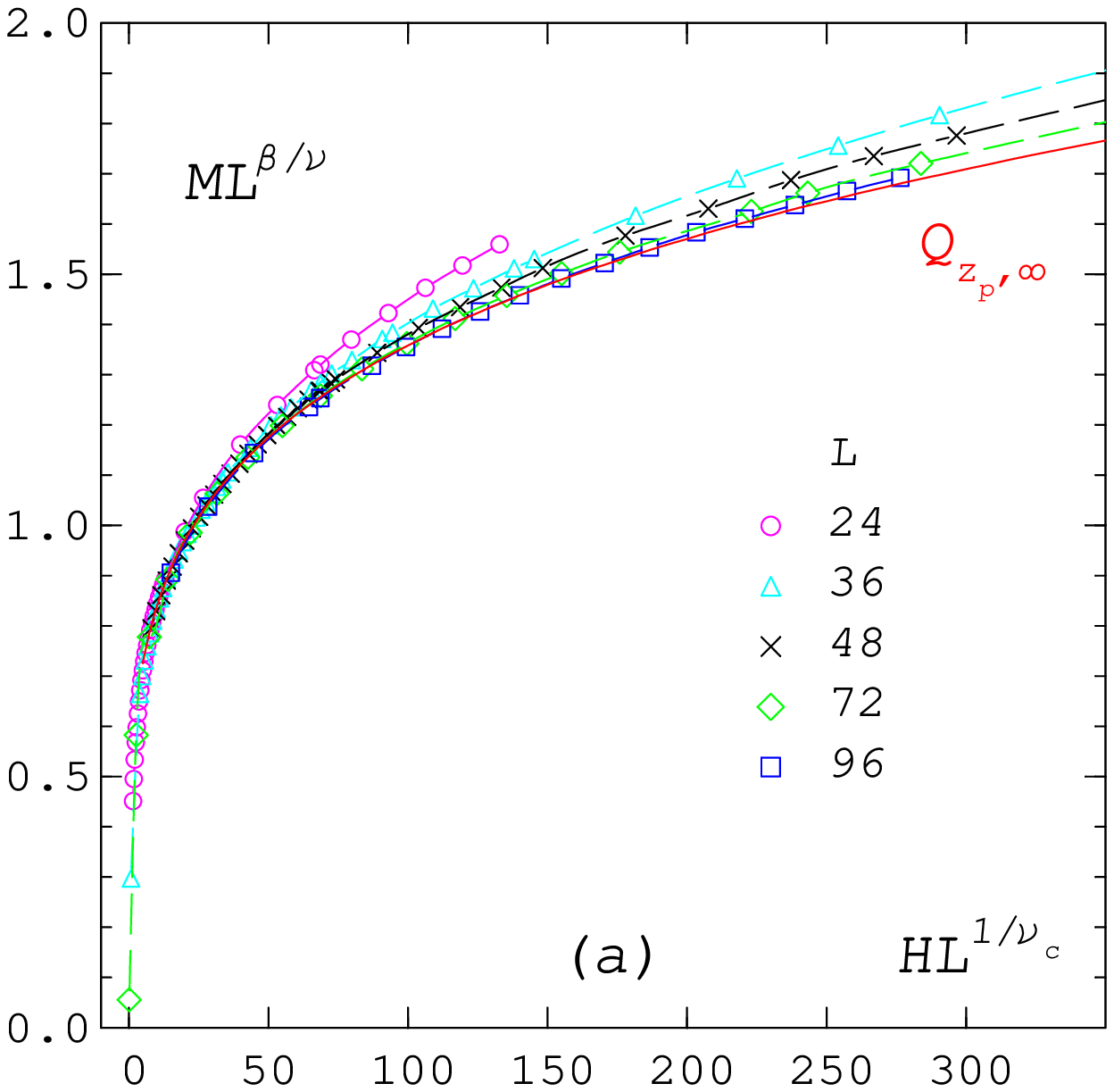, width=67mm}
          }
\put(7.35,0){
   \epsfig{bbllx=127,bblly=264,bburx=451,bbury=587,
       file=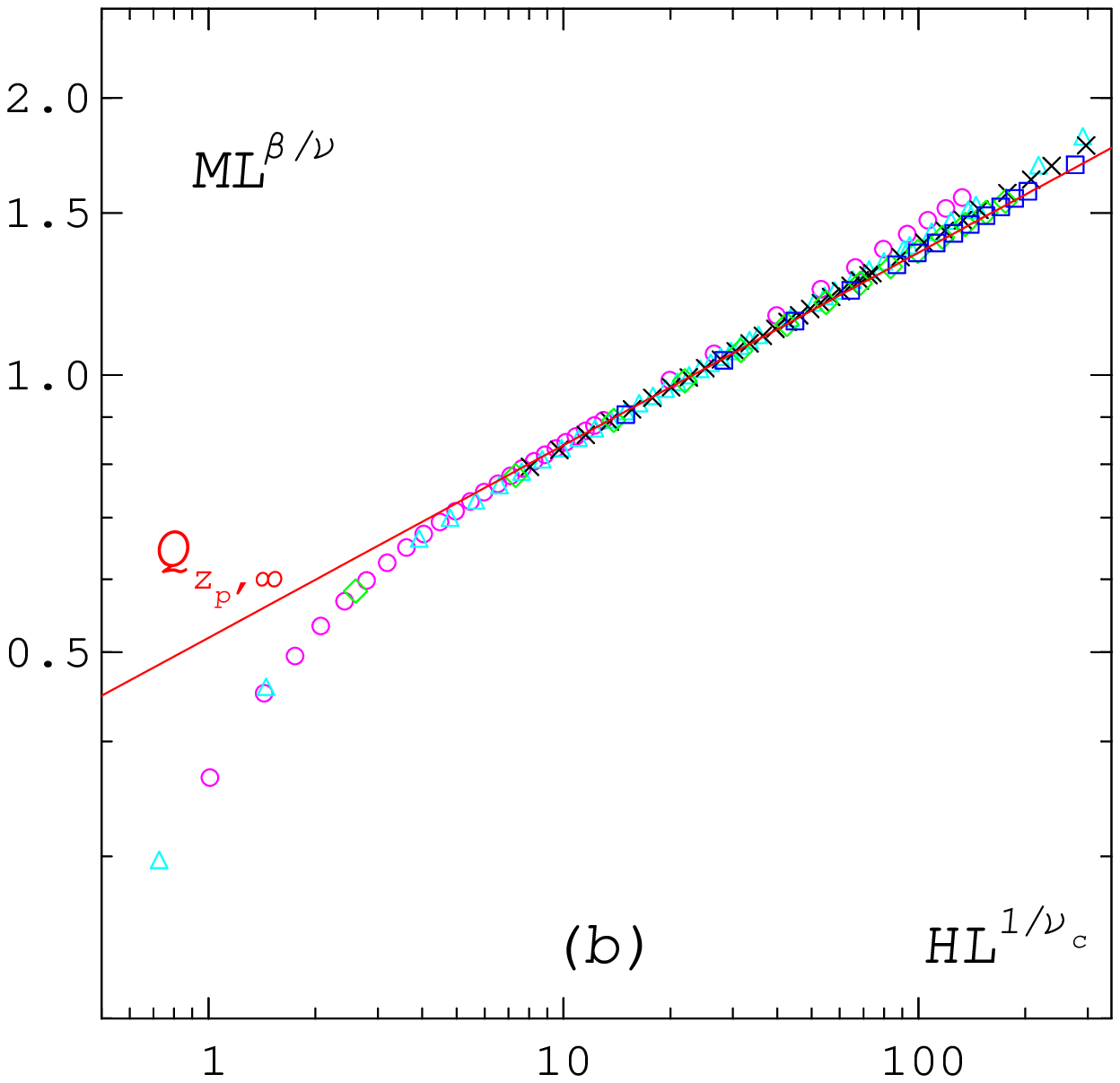, width=67mm}
          }
\end{picture}
\begin{figure}[h!]
\caption{(a) Finite-size scaling of $ML^{\beta/\nu}$ for $O(2)$ 
on the pseudocritical line. The solid line shows the
asymptotic form $Q_{z_p,\infty}$, the symbols denote different lattice
sizes $L$. (b) is a double-log plot of (a).}
\label{fig:o2pc}
\end{figure}

\n $L \ge 24$ in a logarithmic plot. As for the $O(4)$ model we observe an
approach  
of $Q_0$ from below to $Q_{0,\infty}$; from $HL^{1/\nu_c}\approx 10$ on
the two curves coincide, that is $Q_0$ is asymptotic. On the pseudocritical 
line (we have used a somewhat larger value $z_p=1.67$ for $O(2)$ ) we find
again - like for $O(4)$ - an approach of the finite lattice results from
above to the asymptotic finite-size-scaling function as can be seen from
Fig.\ \ref{fig:o2pc}. From the different correction behaviours along the 
critical and pseudocritical lines in both $O(N)$ models one may speculate 
upon the existence of an intermediate $z$ value where the corrections 
disappear. It is unclear, however, what type of corrections to the universal
scaling functions $Q_z$ will be present in QCD.

\section{Comparison to $N_f=2$ QCD}
\label{section:QCD}

We mentioned already in the introduction QCD lattice calculations 
for two light-quark flavours in the staggered formulation 
\cite{JLQCD}-\cite{MILC}. The temperature and the magnetic field which 
one uses in our context here are defined, except for two metric factors, by
\be
t \sim {6 \over g^2} -  {6 \over g_c^2(0)}\quad \mbox{and} \quad 
h \sim m_q aN_{\tau}~.
\label{tandh}
\ee
The coupling $g_c(0)$ denotes the critical coupling in the limit 
$m_q \rightarrow 0$ on a lattice with a fixed number $\NT$ of points in 
the temporal direction. The critical point is that of the chiral transition 
with $\PBP$ as order parameter or magnetization. 
Correspondingly, the pseudocritical coupling $g_c(m_q)$ is given by the 
location of the peak of the chiral susceptibility $\chi_m$ at fixed quark
mass $m_q$. By universality arguments the pseudocritical line is then
predicted as
\be
 {6 \over g_c^2(m_q)} = {6 \over g_c^2(0)} +c m_q^{1/\beta\delta}~.
\label{psline}
\ee 
If the two metric factors normalizing $t$ and $h$ are known, the constant 
$c$ in Eq. (\ref{psline}) is fixed by the universal value of $z_p$.

 In 1998 the JLQCD collaboration \cite{JLQCD} determined the peak heights 
and positions of $\chi_m$ at $m_qa=0.01,0.02,0.0375,0.075$ on lattices 
with spatial sizes $8^3,12^3$ and $16^3$ and $\NT=4$ and found reasonable
agreement with Eq. (\ref{psline}) for $O(4)$ or $O(2)$. We have evaluated
the $\PBP$ data \cite{Kazuy} of the JLQCD collaboration at the peak 
positions listed in Table II of their paper. The resulting values are 
shown in a finite-size-scaling plot with $O(4)$ exponents in Fig.\ 
\ref{fig:scjpbi}a.
On lattices of the same sizes and also at the same quark masses, apart 
from the lowest one, the Bielefeld group \cite{Edqcd} calculated $\PBP$ at 
their own peak positions \cite{Edw}. These data are plotted in Fig.\ 
\ref{fig:scjpbi}b in the same way as those of the JLQCD collaboration. 
In both parts of Fig.\ \ref{fig:scjpbi} we see a behaviour which is similar
to the one in Fig.\ \ref{fig:o4pc}a. The corrections to scaling are such
that the finite-size-scaling function seems to be approached from above. 
Only at smaller values of the scaling variable, that is here for smaller 
values of the quark mass, it appears that the data from all lattices are
higher than expected. This is even more visible in the logarithmic plot,
Fig.\ \ref{fig:comp}, where we show all the data together. Evidently,
instead of falling rapidly at small masses, there is even a relative
increase.
\newpage
%
\setlength{\unitlength}{1cm}
\begin{picture}(13,7.2)
\put(-0.15,0){
   \epsfig{bbllx=127,bblly=264,bburx=451,bbury=587,
       file=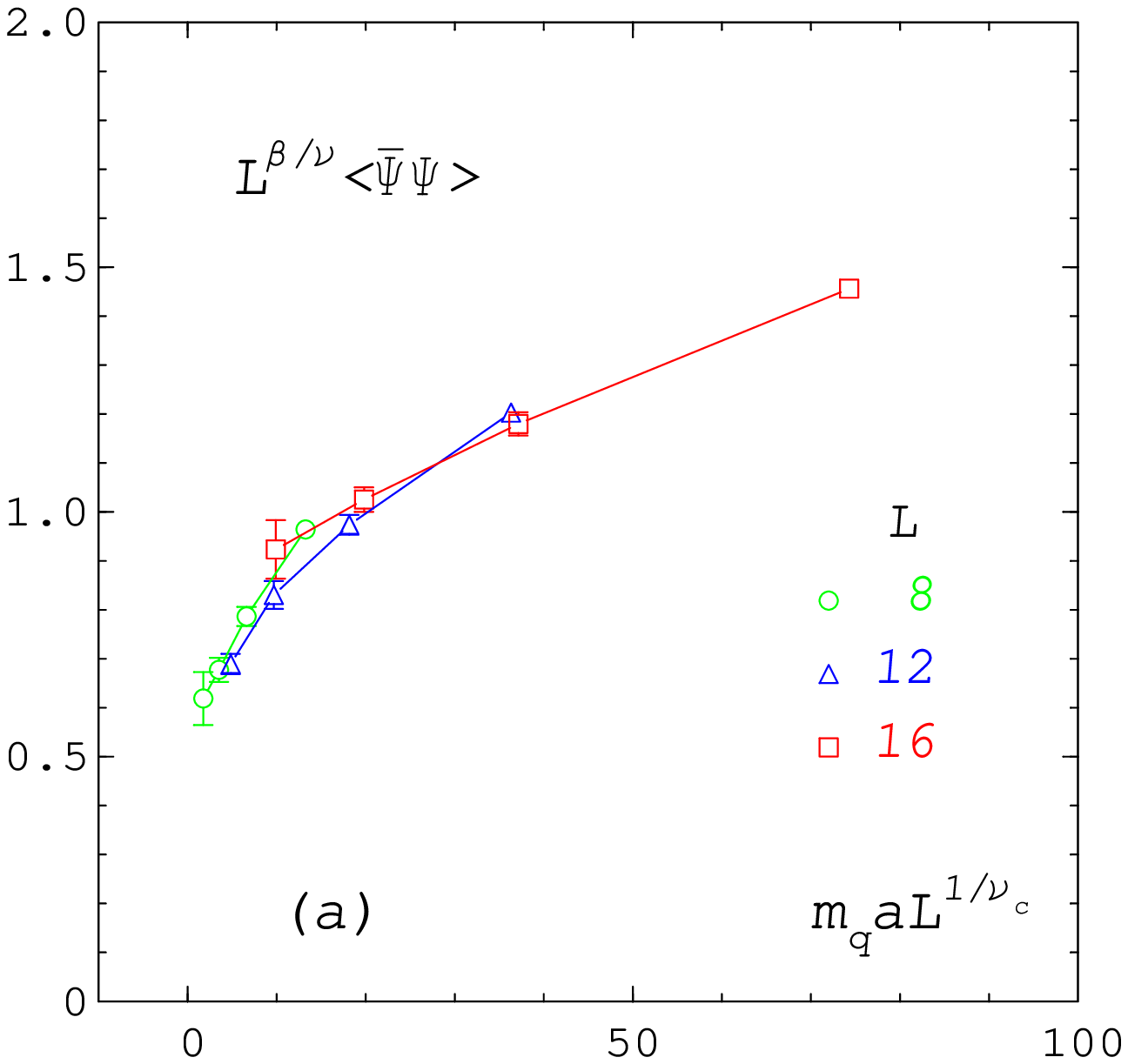, width=67mm}
          }
\put(7.35,0){
   \epsfig{bbllx=127,bblly=264,bburx=451,bbury=587,
       file=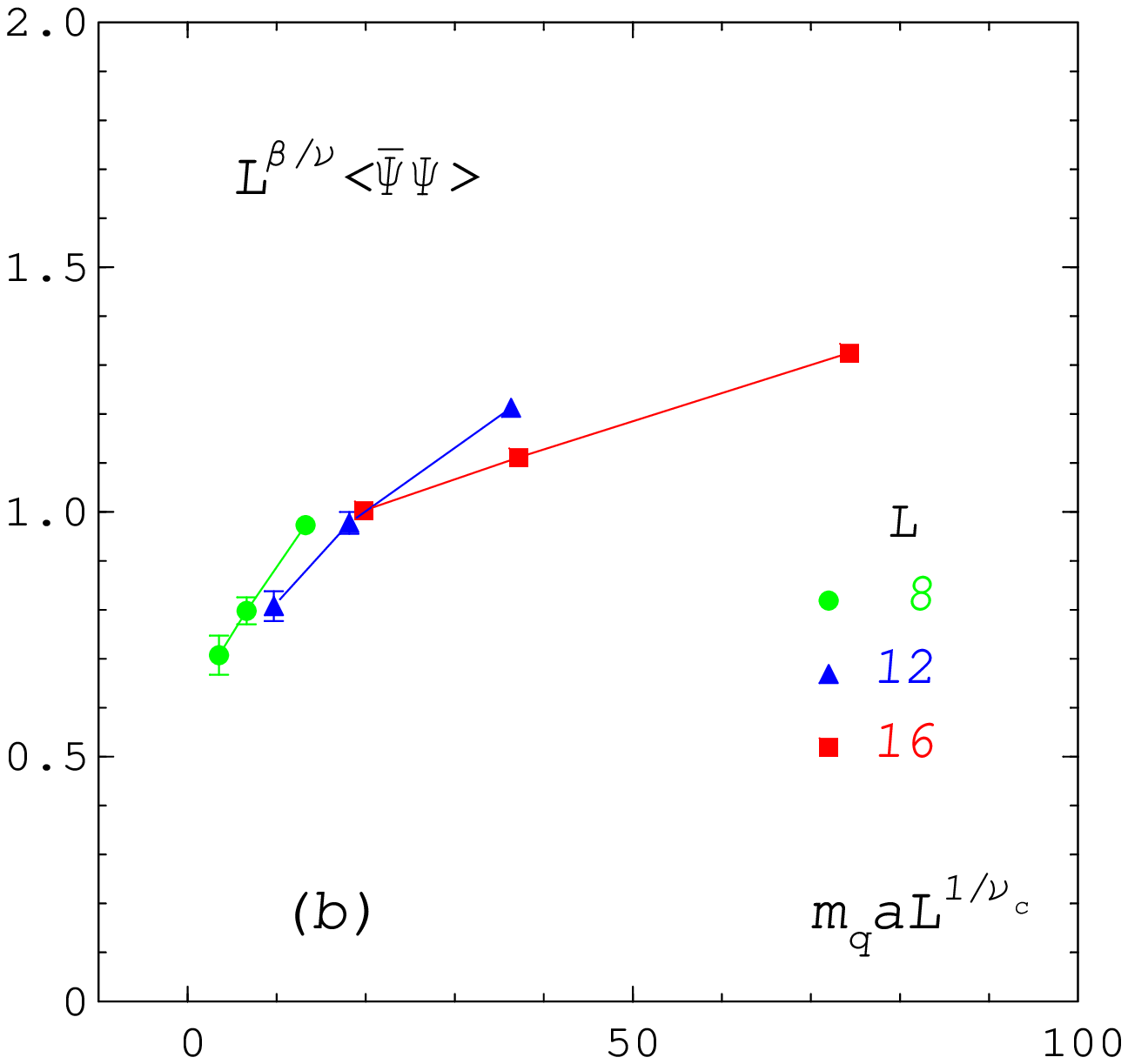, width=67mm}
          }
\end{picture}
\begin{figure}[h!]
\caption{$L^{\beta/\nu}\PBP$ at the peak positions of $\chi_m$ versus
$m_q a L^{1/\nu_c}$ from QCD lattice data with two degenerate staggered 
fermions. The exponents are from the $O(4)$ model.
(a) JLQCD collaboration \cite{Kazuy}, (b) Bielefeld group
\cite{Edw}. The lines are drawn to guide the eye.}
\label{fig:scjpbi}
\end{figure}
\begin{figure}[hb]
\begin{center}
   \epsfig{bbllx=127,bblly=264,bburx=451,bbury=587,
        file=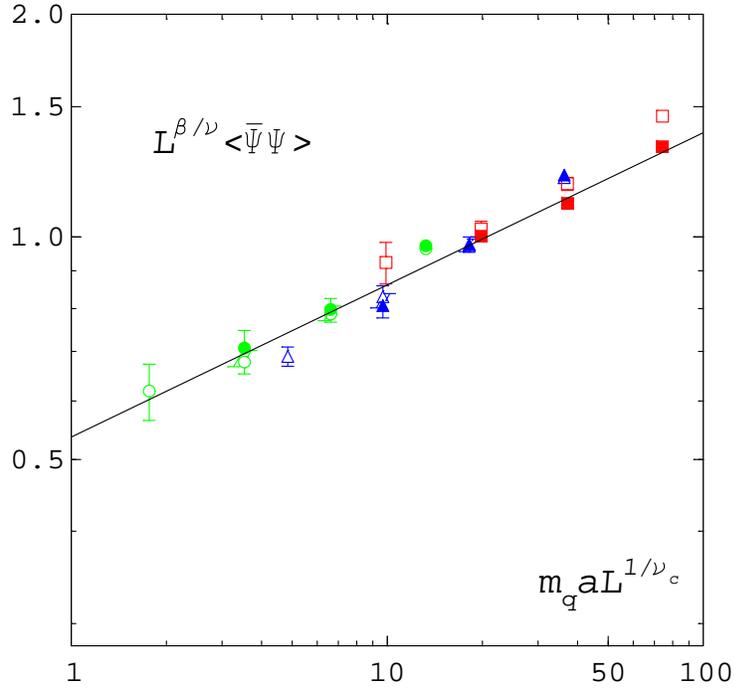,width=84mm}
\end{center}
\caption{Logarithmic plot of Fig.\ \ref{fig:scjpbi}. The QCD data of JLQCD
(empty symbols) and Bielefeld (filled symbols) are compared to the 
asymptotic $O(4)$-finite-size-scaling function (solid line). The notation
for the symbols is as in Fig.\ \ref{fig:scjpbi}.} 
\label{fig:comp}
\end{figure}

\n On the other hand the value of $\PBP$ at very small $m_q$
is more sensitive to the exact position of evaluation, because $\PBP$ is steeper
there. Moreover, we are not precisely on a line of fixed $z$ and there is an
additional finite-size effect due to the separate position determination
at each point and for each lattice size. The error in the position location has
not been taken into account in the plots. With increasing quark mass the data 
points in Fig.\ \ref{fig:comp} must follow a straight line with slope
$1/\delta$, if the universality hypothesis \cite{PW}-\cite{RW} is true.     
We have therefore compared the data to a line $\bar c +(1/\delta)\ln (m_q 
a L^{1/\nu_c})$, which represents the asymptotic finite-size-scaling 
function. Because of the unknown metric factors of QCD the constant $\bar c$ 
was chosen freely. We see in this comparison that the data are indeed
compatible with the expected behaviour, especially when we take into account  
that the lattice sizes are still small and corrections are probably present.
We have repeated the analysis with $O(2)$ exponents. They differ only slightly
from the ones of $O(4)$: $\beta/\nu$ by 1.4\%, $\nu_c$ by 0.3\% and
$\delta$ by 1.7\%. The result is very similar to $O(4)$ and because of the
spread of the data, one cannot really distinguish the two cases. 


\section{Summary and Conclusions}
\label{section:conclusion}

We have investigated finite-size-scaling (FSS) functions for the 
three-dimensional $O(4)$ and $O(2)$ spin models. Our aim was to provide
a more suitable basis for a test of QCD lattice data on the conjectured 
universality class. In order to reduce the number of variables on which
these FSS functions depend, we have calculated these functions along lines
of fixed $z=th^{-1/\beta\delta}$ in the $(t,h)$-plane. This choice was 
motivated by two prominent examples of such lines: the critical line with
$z=0$ and the pseudocritical line of peaks of the susceptibility.
Simulations of QCD are usually performed in the neighbourhood of that line.
In the $O(N)$ models we found the pseudocritical line from the known 
universal scaling functions for $V\rightarrow \infty$. The result was
confirmed by a search with finite volume calculations.

On the critical line we found almost no corrections to scaling for $O(4)$,
while for $O(2)$ strong ones appear, as would be expected. For both models
the universal FSS functions are approached from below with increasing
volume. On the pseudocritical line there are considerable 
corrections to scaling for both models. Here the approach to the universal
FSS functions is from above. In both models and on both lines the 
asymptotic forms of the FSS functions are reached already at small values
around 10 to 30 of the scaling variable $HL^{1/\nu_c}$ from below.   
 
We have made FSS plots from two sets of $N_f=2$ QCD lattice data for $\PBP$
at the peak positions of the susceptibility $\chi_m$. The general behaviour
of the data is similar to that of the $O(N)$ models from finite volumes.
We find an approach to a limiting function from above, though at small 
quark masses (that is at small magnetic fields) the QCD data seem to be too
high. The slope in the logarithmic plot of the data is nevertheless in 
nice agreement with the expectation $1/\delta$ of the $O(N)$ models.
A test on the critical line would be even more preferable, because there 
the $t$ value is independent of $h$. The exact critical point of QCD
is however difficult to determine and up to now unknown.

\vskip 0.2truecm
\noindent{\Large{\bf Acknowledgements}}


\n We owe special thanks to Kazuyuki Kanaya for sending us his complete 
chiral condensate data and to Edwin Laermann for helpful discussions
and his QCD data on the pseudocritical line. We are grateful to David 
Miller for a careful reading of the manuscript. Our work was supported 
by the Deutsche Forschungs\-ge\-meinschaft under Grant No.\ Ka 1198/4-1,
the work of T.M. in addition by FAPESP, Brazil (Project No.00/05047-5).


\clearpage

\end{document}